\documentclass{article}%
\usepackage{booktabs}
\usepackage{array,blindtext}
\usepackage{cite}
\usepackage{hyperref}
\usepackage{amsmath}
\usepackage{amsfonts}
\usepackage{amssymb}
\usepackage{graphicx}%
\DeclareMathOperator{\sgn}{sgn}

\begin{document}

\title{The role of the next-to-leading order triangle-shaped diagram in two-body hadronic decays}
\author{Jonas Schneitzer${}^{\text{a}}$, Thomas Wolkanowski${}^{\text{a}}$ and Francesco Giacosa${}^{\text{a,b}}$\\ \\{\normalsize {\emph{${}^{\text{a}}$Institut f\"{u}r Theoretische Physik,
Goethe-Universit\"{a}t Frankfurt am Main, }}}\\{\normalsize {\emph{60438 Frankfurt am Main, Germany}}}\\{\normalsize {\emph{${}^{\text{b}}$Institute of Physics, Jan Kochanowski University, 25406 Kielce, Poland}}}}
\maketitle

\begin{abstract}
The next-to-leading-order contribution to the amplitude of a two-body decay process is a
triangle-shaped diagram in which the unstable state is exchanged by the
emitted particles. In this work we calculate this diagram in the framework of
a scalar quantum field theory and we estimate its role in hadronic physics, \emph{i.e.}, we apply our results to the well-known scalar-isoscalar resonances $f_{0}(500)$, $f_{0}(980)$, $f_{0}(1370)$,  $f_{0}(1500)$, $f_{0}(1710)$ and the scalar-isovector resonance $a_{0}(1450)$. It turns out that, with the exception of the broad resonance $f_{0}(500)$, the next-to-leading-order contribution is small and can be neglected.
\\
\\
Keywords: two-body decay, scalar resonance \\
PACS numbers: 14.40.Be
\end{abstract}

\section{Introduction}
The study of decays is an important subject of atomic, nuclear and particle
physics \cite{ghirardi}. Some subatomic particles possess a lifetime which is
so short that they can be seen only through their decay products, and hence
one usually calls them resonances. This is indeed the case for the recently
discovered Higgs particle, see \emph{e.g.} Refs. \cite{djouadi,ellis} and references
therein. In the realm of the strong interactions also many hadrons were
discovered via their decay processes \cite{amslerrev}; in addition to that,
decays turn out to be crucially important for the understanding of their quantum
numbers and inner structure.

The main problem concerning the fundamental theory of quarks and gluons
(Quantum Chromodynamics or {\em QCD}) is the fact that this theory is
non-perturbative in the low-energy regime. Hence one relies on other
approaches, as for instance effective models based on symmetries
\cite{geffen,meissner}, where the physical degrees of freedom are not quarks
and gluons, but composite particles, namely hadrons. Decays of hadrons have often
been evaluated within such models in the lowest order approximation -- in
other words at tree-level, see \emph{e.g.} Refs. \cite{ko,tq,elsm,dick} and
references therein. In particular, in the recent work of Ref. \cite{dick}
decays of various mesons up to $1.5$ GeV were computed in a chirally and
dilatation invariant framework and were found to be in agreement with the
corresponding experimental values as provided by the PDG \cite{pdg}.

\begin{figure}[h]
\centering
\hspace{-0.4cm}\includegraphics[scale=0.75]{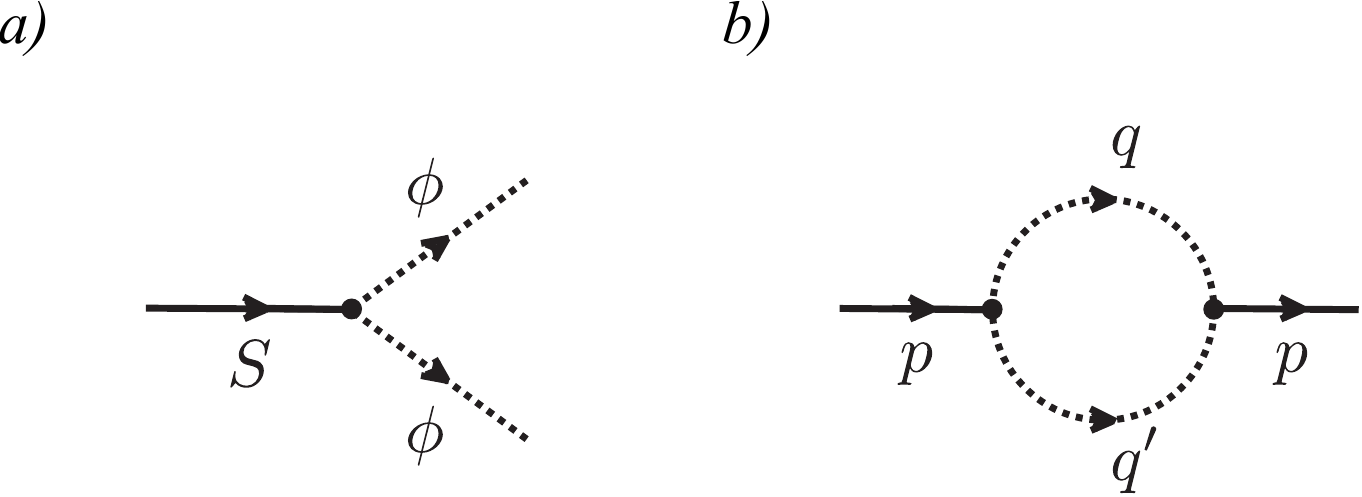}
\caption{a) Decay process $S\rightarrow\phi\phi$ at leading order (tree-level). b) Self-energy at
leading order.}
\label{fig:fig1}
\end{figure}

A two-body tree-level decay is the easiest nontrivial process in quantum field
theory. It is depicted in Fig. \ref{fig:fig1}a: The unstable bosonic
particle $S$ decays into two identical particles, denoted as
$\phi$. The decay amplitude is simply a constant in the case of scalar
particles and non-derivative interactions. When derivatives and/or particles
with nonzero total angular momentum $J$ are considered, a dependence of the
momenta appears in the tree-level amplitude(s).

The next step in the context of effective models has been the study of
(hadronic) loops, see for instance Refs.
\cite{dullemond,coito,tornqvist,pennington,achasov,giacosaSpectral,duecan,e38,salam,Caprini,garcia}
and references therein. The leading contribution to the self-energy is shown
by the diagram in Fig. \ref{fig:fig1}b. Both the mass and the width of the decaying particle are influenced by the quantum fluctuations due
to the coupling to hadronic intermediate states. The optical theorem assures
that the imaginary part of the one-loop diagram from Fig. \ref{fig:fig1}b coincides with the tree-level decay formula. The
unstable particle is described by a spectral function (\emph{i.e.}, an energy
distribution), which is given by the imaginary part of the one-loop resummed
propagator. Alternatively, the properties of the unstable particle can also be
described as a complex pole in the appropriate unphysical Riemann sheet, a
procedure first proposed by Peierls \cite{peierls} a long time ago. The
quantum theoretical treatment of unstable particles became an object of much
interest, see \emph{e.g.} Refs. \cite{hoehler,levypoles,aramaki,landshoff}. The general
outcome of such studies is -- disregarding problems like mixing -- that when
the particle is narrow-shaped, quantum fluctuations have a small influence on
its properties (\emph{i.e.}, mass and width) but are non-negligible for broad
resonances. It turns out to be the ratio `width over mass' that is decisive
here: as long as this number is smaller than $\sim0.3$ the loop contributions
have a small impact \cite{giacosaSpectral}.

\begin{figure}[ptb]
\centering
\hspace{-1.0cm}\includegraphics[scale=0.75]{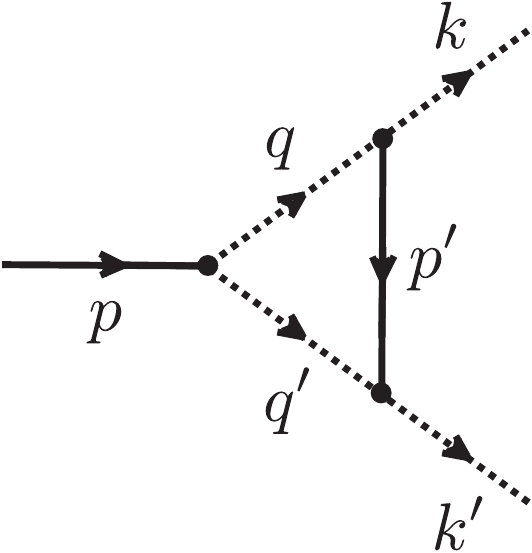}\caption{Triangle-shaped NLO diagram for a two-body decay.}
\label{fig:fig2}
\end{figure}

There is, however, another open issue: what is the role of the next-to-leading
order (NLO) diagram for hadronic decays? We depict this kind of
triangle-shaped diagram in Fig. \ref{fig:fig2}; it is proportional
to the third power of the coupling constant. In the context of hadronic decays
in effective field theories/models it is usually not taken into account.
Nevertheless, one should stress that the coupling constant in hadronic models
is in general not a small number, thus there is a priori no guarantee that the
NLO diagram is smaller than the tree-level one.

The aim of this work is to close this gap. To this end, we evaluate the role
of the triangle diagram from Fig. \ref{fig:fig2} in the case of a
simple scalar field theory without derivative interactions. We do this in
plain perturbation theory (\emph{i.e.}, without resummation), meaning that the
virtual $S$-particle exchanged in Fig. \ref{fig:fig2} is described by
its free propagator. After discussing the analytic properties of the triangle
diagram, we adopt our results to some decays of well-known scalar resonances:
$f_{0}(500)$, $f_{0}(980)$, $f_{0}(1370)$ and $f_{0}(1500)$ \cite{pdg}. All
these resonances decay predominately in two pions and are therefore a good
test for our purpose. For completeness, we also look at the pion-pion and kaon-kaon decay channels of $f_{0}(1710)$ as well as the kaon-kaon decay of the scalar-isovector state $a_{0}(1450)$ \cite{pdg}. Yet, our investigation is quite general and applies also to decays involving derivatives and particles with spin.

The main result of our work is that the triangle contribution is indeed
\emph{negligible} and one can consequently justify a posteriori all previous
studies in which those types of contributions (and, in turn, higher-order
contributions as well) were not taken into account. Since in the field of
hadron physics there are usually other (and even larger) sources of
uncertainties due to various (and sometimes subtle) approximations and
simplifications, the restriction to the leading order tree-level diagram from
Fig. \ref{fig:fig1}a and to the (resummed) one-loop quantum
corrections from Fig. \ref{fig:fig1}b are reasonable and usually sufficient.

The paper is organized as follows: we present the model and some
analytic aspects of the triangle diagram in Section \ref{seq:model}, while the numerical
results are shown in Section \ref{seq:results}. The last Section \ref{seq:conc} contains the conclusions.

\section{The model and the triangle diagram}
\label{seq:model}
We introduce a model with the scalar fields $S$ (with mass $m_{S}$) and $\phi$ (with mass $m_{\phi}$) described by the Lagrangian
\begin{equation}
\mathcal{L}=\frac{1}{2}(\partial_{\mu}S)^{2}-\frac{m_{S}^{2}}{2}S^{2}+\frac{1}{2}(\partial_{\mu}\phi)^{2}-\frac{m_{\phi}^{2}}{2}\phi^{2}+gS\phi^{2} \ .
\end{equation}
The interaction term induces the decay process $S\rightarrow\phi\phi$. The
parameter $g$ is the coupling constant (with dimension of energy). For
previous studies and details of the model see Refs.
\cite{achasov,giacosaSpectral,duecan,e38,veltman,thomasthesis}. The decay
width in perturbation theory can be expressed as
\begin{equation}
\Gamma_{S\rightarrow\phi\phi}=\frac{1}{2}\frac{\sqrt{\frac{m_{S}^{2}}%
{4}-m_{\phi}^{2}}}{8\pi m_{S}^{2}}|\hspace{0.02cm}\mathbf{-}\hspace
{0.05cm}i\mathcal{M}|^{2}\text{ } \ , \label{gammafull}
\end{equation}
where the decay amplitude $-i\mathcal{M}$ in perturbation theory is written as
a sum
\begin{equation}
-i\mathcal{M}=-i\mathcal{M}_{1}-i\mathcal{M}_{3}+\dots \ .
\end{equation}
The term $-i\mathcal{M}_{2n+1}$ represents the contribution of order
$g^{2n+1}$ to the amplitude. It is quite remarkable that an exact solution to
this problem has not yet been found (there are, however, quantum mechanical
models for which this is possible, see Refs.
\cite{ghirardi,lee,kofman,facchiprl,giacosapra,duecan,shimizu}).

\bigskip

1) \emph{Leading order}: The leading order with $n=0$ is given by the
tree-level amplitude of Fig. \ref{fig:fig1}a, for which one obtains
\begin{equation}
-i\mathcal{M}_{1}=2ig \ .
\label{eq:tree_amplitude_phiphi}
\end{equation}
Thus, the tree-level decay width simply reads
\begin{equation}
\Gamma_{S\rightarrow\phi\phi}^{\text{tree}}=\frac{1}{2}\frac{\sqrt{\frac
{m_{S}^{2}}{4}-m_{\phi}^{2}}}{8\pi m_{S}^{2}}\hspace{0.03cm}(2g)^{2} \ .
\label{eq:tree_width_phiphi}
\end{equation}

2) \emph{The one-loop diagram}: The loop diagram shown in Fig.
\ref{fig:fig1}b actually does \emph{not} enter directly into the
expression of Eq. (\ref{gammafull}), because the latter is valid in plain
perturbation theory without resummation. The role of the one-loop resummation
has been widely studied and, although not directly relevant for our
calculation, we recall the main features in view of its general importance for
the problem that we are studying and for future works \cite{dullemond,coito,tornqvist,pennington,achasov,giacosaSpectral,duecan,e38,thomasthesis}%
. By denoting $\Pi(p^{2})$ as the loop contribution, the propagator of the
field $S$ changes upon resummation to
\begin{equation}
\Delta_{S}(p^{2})=\frac{i}{p^{2}-m_{S}^{2}+i\epsilon}\ \ \rightarrow \
\ G_{S}(p^{2})=\frac{i}{p^{2}-m_{S}^{2}-\Pi(p^{2})+i\epsilon} \ .
\end{equation}
The self-energy $\Pi(p^{2})$ is linked to the tree-level decay width via the
optical theorem:
\begin{equation}
-\operatorname{Im}\Pi(p^{2})=\sqrt{p^{2}}\ \Gamma_{S\rightarrow\phi\phi
}^{\text{tree}}\hspace{0.03cm}(m_{S}^{2}\rightarrow p^{2})=\sqrt{p^{2}}\ \frac{g^{2}\sqrt{\frac{p^{2}}{4}-m_{\phi}^{2}}}{4\pi p^{2}} \ .
\end{equation}
The properties of the unstable particle $S$ (\emph{i.e.}, its mass and decay width)
are often identified with the complex pole of the full propagator $G_{S}%
(p^{2})$ in the second Riemann sheet, $p^{2}=\big(m_{S}^{\text{pole}}%
-i\hspace{0.03cm}\Gamma_{S\rightarrow\phi\phi}^{\text{pole}}/2\big)^{2}$. The
(shifted) mass $m_{S}^{\text{pole}}$ is given by the real part of the pole and
the modified decay width $\Gamma_{S\rightarrow\phi\phi}^{\text{pole}}$ by the
negative imaginary part multiplied by two. For a small coupling constant $g$ the
quantities $\Gamma_{S\rightarrow\phi\phi}^{\text{pole}}$ and $\Gamma
_{S\rightarrow\phi\phi}^{\text{tree}}$ are close to each other, but then
deviate when increasing $g$ \cite{thomasthesis}.

At this point it should be stressed that hadronic theories do not need to
undergo a renormalization process as for theories of elementary particles,
since they are only valid in a limited energy regime. (For instance, if we
restrict our attention to mesons made of $u$-, $d$-, and $s$-quarks, the
associated `cutoff' has a value of about $\sim1.5$ GeV.) For this reason, a
finite energy cutoff is used when evaluating the loop diagram from Fig.
\ref{fig:fig1}b. Different forms for the consequently needed cutoff
function can be applied, but the general outcome shows just a soft dependence on
the precise choice as long as convergence is guaranteed \cite{giacosaSpectral}.

In an ideal scattering experiment of the type $\phi\phi\rightarrow\phi\phi,$
the unstable state $S$ manifests itself as an enhanced peak for a center of
mass energy close to $m_{S}.$ More precisely, the spectral function
$d_{S}(E)=-\frac{2E}{\pi}\operatorname{Im}G_{S}(E^{2})$ plays an important
role: as was argued long time ago by Matthews and Salam \cite{salam}, it can
be interpreted as a `mass distribution' of the unstable particle $S$, which
can be well described by a Breit--Wigner function for narrow resonances.
Namely, even if there are low-energy threshold(s) and high-energy distortions,
as long as the ratio $\Gamma_{S\rightarrow\phi\phi}^{\text{tree}}/m_{S}$ is a
small number, the role of hadronic loop contributions is small
\cite{giacosaSpectral,duecan}. However, when this ratio becomes large one
observes big deviations from a typical Breit--Wigner peak and in some cases a
very peculiar phenomenon takes place, often called pole-doubling. This means
that new poles can emerge in the unphysical Riemann sheet(s)
\cite{tornqvist,pennington,dullemond}. Although there is only one unstable
state in the Lagrangian, quantum fluctuations might be able to generate two
(or more) resonance poles. Such a
mechanism of dynamical generation is present in the ongoing debate among
hadron physicists on where some of the (known) resonances in the hadron
spectrum arise from.\footnote{For a general discussion concerning dynamical
generation see Ref. \cite{giacosaDynamical} and references therein.} A
particular interesting field of study is that of charmed mesons, see Refs.
\cite{coito,brambilla}.

\newpage

3) \emph{The triangle diagram}: We now turn our attention to the main subject of this work: the NLO (non-resummed) perturbation theory. To this end, we evaluate the
triangle diagram corresponding to the amplitude $-i\mathcal{M}_{3}$ as
depicted in Fig. \ref{fig:fig2}. Its analytic expression takes the
form
\begin{align}
-i\mathcal{M}_{3}(p  &  \rightarrow kk^{\prime})\label{eq:3amplitude}\\
&  =\ 8ig^{3}\int\frac{\text{d}^{4}q}{(2\pi)^{4}}\ \frac{1}{q^{2}-m_{\phi}^{2}+i\epsilon}\hspace{0.03cm}\frac{1}{(q-p)^{2}-m_{\phi}^{2}+i\epsilon}\hspace{0.03cm}\frac{1}{(q-k)^{2}-m_{S}^{2}+i\epsilon} \ . \nonumber
\end{align}
Note, the propagator of $S$ is taken as the free propagator $\Delta_{S}%
(p^{2}).$ The factor of 8 arises due to identical particles in the final state. Solving the integration over $q^{0}$ by the residue theorem and
after introducing spherical coordinates the amplitude can be re-expressed
as
\begin{equation}
\mathcal{M}_{3}(p\rightarrow kk^{\prime})=\frac{2g^{3}}{\pi^{2}}\int
_{0}^{\infty}\mbox{d}u\;u^{2}\int_{-1}^{1}\mbox{d}\chi\;\big\{P_{12}(u,\chi)+P_{3}(u,\chi)\big\} \ , \label{eq:m3second}
\end{equation}
where
\begin{align}
P_{12}(u,\chi)\ =  &  \ \ \frac{1}{8\sqrt{u^{2}+m_{\phi}^{2}}\big(m_{\phi}%
^{2}+w^{2}(1-\chi^{2})\big)}\\
&  \ \times\ \frac{u^{2}-2w^{2}+uw\chi}{(u-p_{1}-i\epsilon)(u-p_{2}%
+i\epsilon)(u-p_{3}(\chi)-i\epsilon\xi)(u-p_{4}(\chi)+i\epsilon\xi
)} \ ,\nonumber\\[10pt]
\xi\ =  &  \ \hspace{0.08cm}\sgn(-\triangle_{2}-2uw\chi) \ ,\\[7pt]
P_{3}(u,\chi)\ =  &  \ \ \frac{1}{8u\sqrt{u^{2}+m_{S}^{2}}\big(w^{2}(\chi
^{2}-1)-m^{2}\big)\big(u-p_{5}(u,\chi)-i\epsilon\xi^{\prime}\big)} \ ,\\[7pt]
\xi^{\prime}\ =  &  \ \hspace{0.08cm}\sgn\left(\frac{m_{S}^{2}}{u}%
-2w\chi\right)  \ ,\\[7pt]
\triangle_{1}\ =  &  \ \ m_{S}^{2}-m_{\phi}^{2} \ ,\\[5pt]
\triangle_{2}\ =  &  \ \ 2m_{\phi}^{2}-m_{S}^{2} \ .
\end{align}
Here we introduced $|\mathbf{q}|=u$, $|\mathbf{k}|=w$ and $\chi=\cos\theta$,
where $\theta$ is the angle between $\mathbf{q}$ and $\mathbf{k}$. We applied
the shift $\mathbf{q}\rightarrow\mathbf{q}-\mathbf{k}$ for $P_{3}$ before the
transformation to spherical coordinates. The integrand in Eq.
(\ref{eq:m3second}) has five poles:
\begin{align}
p_{1}\ =  &  \ \ w\ ,\label{eq:Pole4Pre}\\
p_{2}\ =  &  \ -\hspace{-0.07cm}w\ ,\\
p_{3}(\chi)\ =  &  \ \ \frac{-w\chi\triangle_{2}-\sqrt{u^{2}\chi^{2}%
\triangle_{2}^{2}-4w^{2}\triangle_{1}(-w^{2}-m_{\phi}^{2}+w^{2}\chi^{2})}%
}{2(-w^{2}-m_{\phi}^{2}+w^{2}\chi^{2})}\nonumber\\[7pt]
\ =  &  \ \ \frac{-2w\chi\triangle_{2}-wm_{S}\sqrt{4w^{2}+m_{\phi}^{2}%
(3+\chi^{2})}}{2(-w^{2}-m_{\phi}^{2}+w^{2}\chi^{2})}\ ,\\[7pt]
p_{4}(\chi)\ =  &  \ \ \frac{-w\chi\triangle_{2}+\sqrt{w^{2}\chi^{2}%
\triangle_{2}^{2}-4w^{2}\triangle_{1}(-w^{2}-m_{\phi}^{2}+w^{2}\chi^{2})}%
}{2(-w^{2}-m_{\phi}^{2}+w^{2}\chi^{2})}\nonumber\\[7pt]
\ =  &  \ \ \frac{-2w\chi\triangle_{2}+wm_{S}\sqrt{4w^{2}+m_{\phi}^{2}%
(3+\chi^{2})}}{2(-w^{2}-m_{\phi}^{2}+w^{2}\chi^{2})}\ ,\\[5pt]
p_{5}(\chi)\ =  &  \ \ \frac{m_{S}^{2}w\chi}{w^{2}(\chi^{2}-1)-m^{2}}\ .
\end{align}
One finds that only $p_{1},$ $p_{3}$ and $p_{5}$ are positive and located on
the path of integration (where, for the latter, this in fact depends on the
value of $\chi$), and thus contribute to the imaginary part of the integral in Eq. (\ref{eq:m3second}). The contribution from $p_{1}$ is
easy to calculate analytically, whereas the one from $p_{3}$ yields a rather
complicated result and is therefore computed numerically. Notice that just the
term $P_{12}$ contains those two singularities and that the contribution of
$p_{5}$ vanishes.

Before moving to our final results, two comments are in order:

\begin{itemize}
\item The NLO diagram in Fig. \ref{fig:fig2} is convergent and
well-defined also for an infinite cutoff. However, as we explained above, a
finite value of the cutoff naturally comes into a hadronic theory because of
the non-elementary nature of the fields
\cite{closecutoff,giacosaSpectral,duecan} and the finite size of the
corresponding particles \cite{tornqvist,pennington}, respectively. One should
in principle evaluate this diagram by including such a cutoff even if
convergence is ensured in the limit of $\Lambda\rightarrow\infty$. It turns
out that for what concerns the triangle diagram the influence of the cutoff
parameter is small -- a value of about $\sim1$ GeV or taking the
infinity limit generates only small changes.

\item Triangle-shaped diagrams were indeed studied in hadron physics, but in a
rather different framework. For instance, one has studied the processes
$\pi^{0}\rightarrow\gamma\gamma$ and $f_{0}\rightarrow\gamma\gamma$
\cite{trianglegammagamma}, where $f_{0}$ represents a generic scalar state
(see Section \ref{seq:results}). These decays occur via a triangle-loop of quarks
and represent the leading order contributions (there is no tree-level diagram
for those processes). For a case in which mesonic loops contribute to $\gamma\gamma$
emission see Ref. \cite{hanhart}. The mentioned investigations have similar technical aspects to our
present interest, yet they could not give an answer to the question of the
role of the triangle diagram as a next-to-leading order contribution.
\end{itemize}

\section{Results}
\label{seq:results}
\subsection{General case}
We first present numerical results of our calculations without referring to any
particular mesonic state. To this end, we fix the energy units in the
following way:
\begin{equation}
\lbrack g]=[m_{\phi}]=[\mathcal{M}]=[\Gamma]=[\Lambda]=1\ m_{S} \ ,
\end{equation}
\emph{i.e.}, all dimensionful quantities are expressed in terms of the mass of the
unstable state $S.$

\begin{figure}
\centering
\hspace{1.0cm}\includegraphics[scale=0.6]{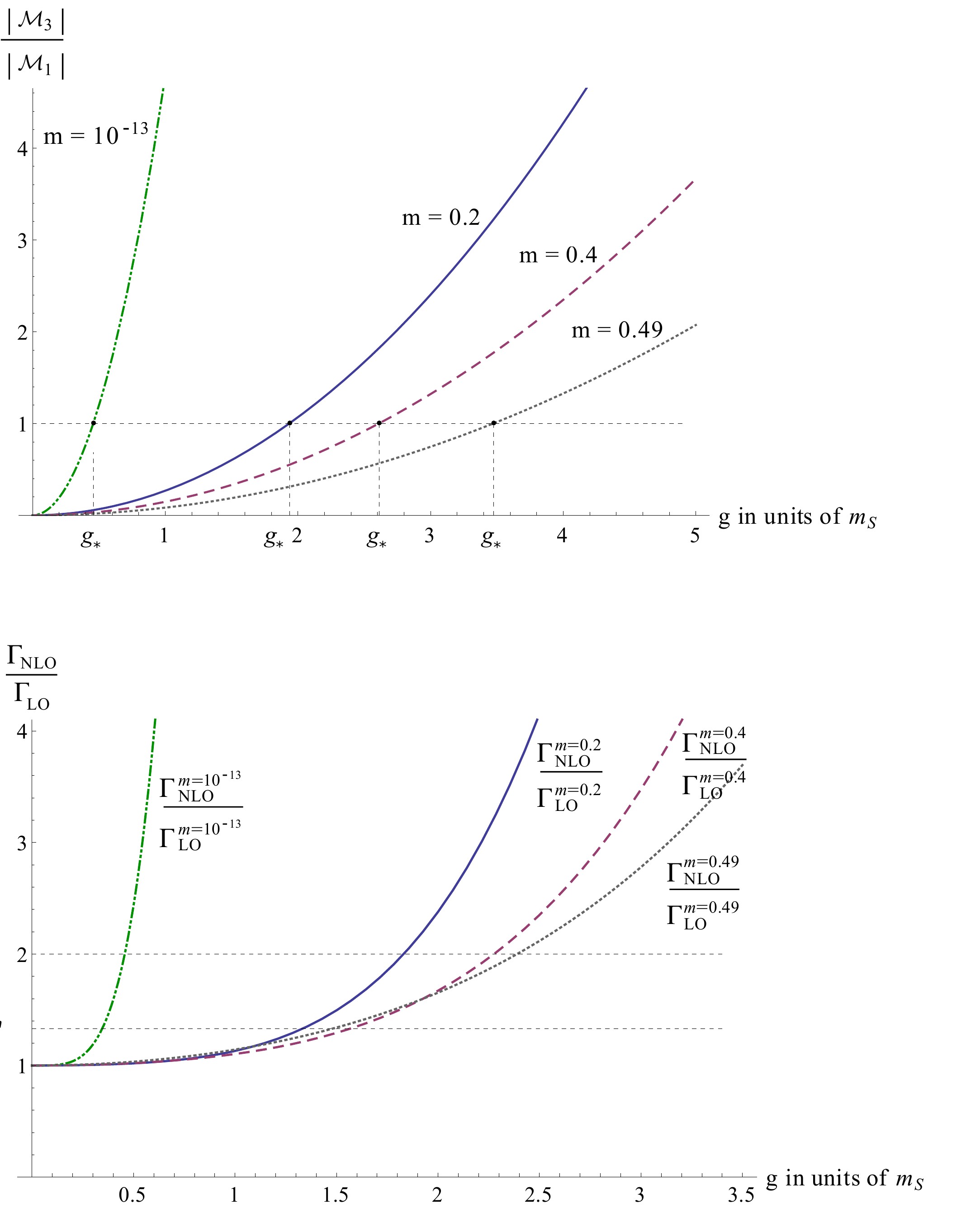}
\caption{Upper panel: Ratio $|\mathcal{M}_{3}|/|\mathcal{M}_{1}|$ of the decay
amplitudes for the mass configurations $m_{\phi}=m=10^{-13},\hspace
{0.05cm}0.1,\hspace{0.05cm}0.2,\hspace{0.05cm}0.3,\hspace{0.05cm}%
0.4,\hspace{0.05cm}0.49$ in dependence of the coupling constant $g$. Lower
panel: Ratio $\Gamma_{\text{NLO}}/\Gamma_{\text{LO}}$ of the decay widths for
the same mass configurations in dependence of the coupling constant $g$.}
\label{fig:fig3}
\end{figure}

In the upper panel of Fig. \ref{fig:fig3} the ratio
$|\mathcal{M}_{3}|/|\mathcal{M}_{1}|$ of the decay amplitudes is shown for
different masses of the particle $\phi$ in dependence of the coupling constant
$g$. In this way the role of the triangle diagram is visualized. As expected,
the larger the coupling, the larger $|\mathcal{M}_{3}|/|\mathcal{M}_{1}|$ and
for equal $g$ the ratio is larger for smaller masses $m_{\phi}=m$. We denote
$g=g_{\ast}$ as the value of the coupling for which the amplitudes are equal,
$|\mathcal{M}_{3}|$ $=|\mathcal{M}_{1}|$, implying that then the triangle
diagram is exactly as large as the tree-level one. The value $g_{\ast}$ represents an upper limit for the validity of a tree-level calculation in particular and for a perturbative expansion in general (see Tab. \ref{tab:results} for a list of numerical values).

In Fig. \ref{fig:fig3}, lower panel, we also show the ratio
$\Gamma_{\text{NLO}}/\Gamma_{\text{LO}}$ for different values of $m_{\phi}$ as
function of $g,$ where the lowest order is the tree-level decay width
$\Gamma_{\text{LO}}=\Gamma_{S\rightarrow\phi\phi}^{\text{tree}}$, meaning $-i\mathcal{M}=-i\mathcal{M}_{1}=2ig$ in Eq. (\ref{gammafull}), and the
next-to-leading order is $\Gamma_{\text{NLO}}$, meaning $-i\mathcal{M}=-i\mathcal{M}_{1}-i\mathcal{M}_{3}$. We also define the two
specific couplings $g^{\prime}$ and $g^{\prime\prime}$, where $g^{\prime}$
corresponds to the ratio $\Gamma_{\text{NLO}}/\Gamma_{\text{LO}}=1.33$ (the decay
width at NLO is $33\%$ larger than the tree-level width, \emph{i.e.}, this marks a `soft' limit for the validity of the tree-level calculation) and $g^{\prime\prime}$ corresponds to $\Gamma_{\text{NLO}%
}/\Gamma_{\text{LO}}=2$ (a `hard' limit). The values of $g^{\prime}$ and
$g^{\prime\prime}$ are reported in Tab. \ref{tab:results}.

\begin{table}
\begin{center}
\begin{tabular}
[c]{|l|c|c|c|}\hline
$m_{\phi}$ & $g_{\ast}$ & $g^{\prime}$ & $g^{\prime\prime}$\\
\hline
\hline
$10^{-13}$ & $0.4581$ & $0.3459$ & $0.4571$%
\\\hline
$0.1$ & $1.6407$ & $1.1766$ & $1.5894$\\\hline
$0.2$ & $1.9365$ & $1.3327$ & $1.8316$\\\hline
$0.3$ & $2.2155$ & $1.4279$ & $2.0164$\\\hline
$0.4$ & $2.6122$ & $1.5672$ & $2.2767$\\\hline
$0.49$ & $3.4742$ & $1.4796$ & $2.3908$\\\hline
\end{tabular}
\end{center}
\caption{Specific coupling constants for different mass configurations: The case $|\mathcal{M}_{3}|$
$=|\mathcal{M}_{1}|$ is obtained for $g=g_{\ast}$, the case $g=g^{\prime}$ for
$\Gamma_{\text{NLO}}/\Gamma_{\text{LO}}=1.33$ and the case $g=g^{\prime\prime}$ for
$\Gamma_{\text{NLO}}/\Gamma_{\text{LO}}=2$. All quantities are in units of $m_{S}$.}
\label{tab:results}
\end{table}

The following comments are in order:

\vspace{0.3cm}
(i) Although the limit $m_{\phi}\rightarrow0$ is subtle because it contains
infrared divergences, even for the very small value $m_{\phi}=m=10^{-13}$ the
NLO becomes dominant only for $g\gtrsim0.45$. This case is however unrealistic
for hadronic physics in which $m_{S}\sim1$ GeV and $m_{\phi}\ge m_{\pi}$.

\vspace{0.3cm}
(ii) For the ratio $m_{\phi}/m_{S}\gtrsim0.1$, which is usually fulfilled for
decays of hadrons, one has $g_{\ast}\sim g^{\prime\prime}\gtrsim1.6$ and
$g^{\prime}\gtrsim1.2$, see Tab. \ref{tab:results}. These values of coupling constants
correspond to very large decay widths. In physical cases the value of $g$ is
usually safely smaller, showing that the NLO is subdominant, see next Subsection for explicit examples.

(iii) The fact that $g_{\ast}\sim g^{\prime\prime}$ is possible because the
NLO amplitude, $-i\mathcal{M}_{3}$, has a dominant imaginary part. As a
consequence
\begin{align}
\Gamma_{\text{NLO}} & = \vspace{0.25cm} \frac{1}{2}\frac{\sqrt
{\frac{m_{S}^{2}}{4}-m_{\phi}^{2}}}{8\pi m_{S}^{2}}|\hspace{0.02cm}%
\mathbf{-}\hspace{0.05cm}i\mathcal{M}_{1}\mathbf{-}\hspace{0.05cm}%
i\mathcal{M}_{3}|^{2}\simeq\vspace{0.25cm}\frac{1}{2}\frac{\sqrt{\frac{m_{S}^{2}}{4}-m_{\phi
}^{2}}}{8\pi m_{S}^{2}}\left(|\hspace{0.02cm}\mathbf{-}\hspace
{0.05cm}i\mathcal{M}_{1}|^{2}+|\hspace{0.02cm}\mathbf{-}\hspace{0.05cm}%
i\mathcal{M}_{3}|^{2}\right) \nonumber \\
& =\vspace{0.25cm}\Gamma_{\text{LO}}+\frac{1}{2}\frac{\sqrt
{\frac{m_{S}^{2}}{4}-m_{\phi}^{2}}}{8\pi m_{S}^{2}}|\hspace{0.02cm}%
\mathbf{-}\hspace{0.05cm}i\mathcal{M}_{3}|^{2} \ ,
\end{align}
which shows that $\Gamma_{\text{NLO}}$ is always larger
than $\Gamma_{\text{LO}}$ (interference effects that
involve only the real part of $\mathbf{-}\hspace{0.05cm}i\mathcal{M}_{3}$ are
small) and that, when the NLO equals the LO, the width doubles.

\vspace{0.3cm}
(iv) Although in our study one has a decay into particles with equal masses,
nothing substantial would change for the decay into two particles with different masses.
The numerical evaluation would be more involved because other poles could contribute.

\subsection{Specific examples}
\begin{figure}
\hspace{-3.0cm}\includegraphics[scale=0.9]{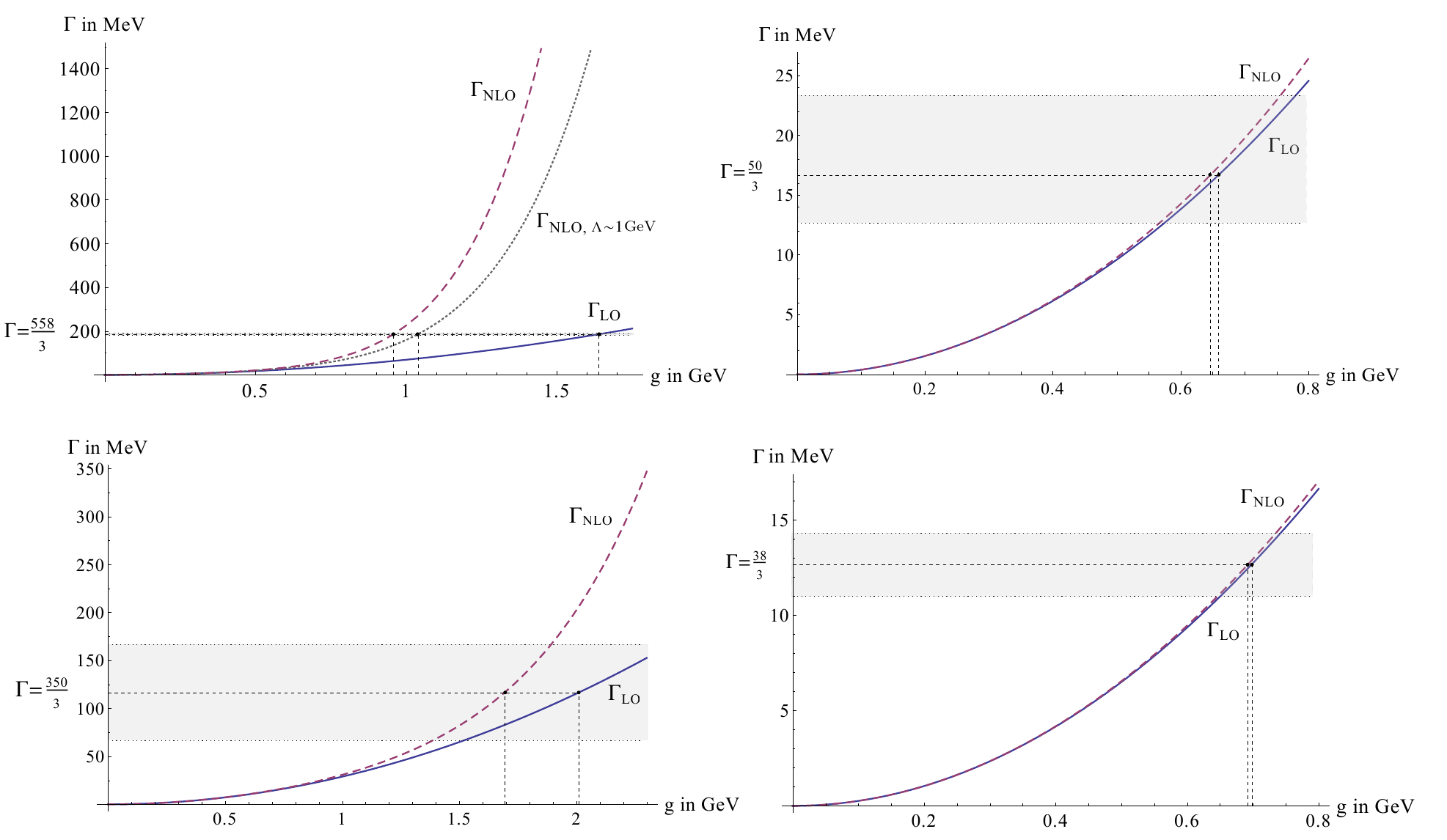}
\caption{Decay widths of $f_{0}(500)$ (upper left), $f_{0}(980)$ (upper
right), $f_{0}(1370)$ (lower left) and $f_{0}(1500)$ (lower right). The
`known' value (for the two resonances below $1$ GeV taken from Ref.
\cite{garcia}, for the other two from the PDG \cite{pdg}) of the width is
marked on the $y$-axis and error bars are indicated by a gray band. Note that here $\Gamma=\Gamma_{f_{0}\rightarrow\pi^{0}\pi^{0}}=\Gamma_{f_{0}\rightarrow\pi\pi}/3$. The vertical lines correspond to the determination of the coupling constant at LO and NLO, respectively.}
\label{fig:fig4}
\end{figure}

We now turn to the concrete examples of the well-known scalar resonances
$f_{0}(500)$, $f_{0}(980)$, $f_{0}(1370)$ and $f_{0}(1500)$ and calculate
their decays into pions \cite{pdg} (for a
discussion of the internal structure of these states in terms of quarks and
gluons see also \emph{e.g.} Refs. \cite{ko,tq,elsm,dick,giacosaSpectral,dullemond,Caprini,lohse,Black}). We present the results in Fig. \ref{fig:fig4}, plotting the decay
widths as function of $g$. The experimental value
of the width (which is $\Gamma_{f_{0}\rightarrow\pi\pi}/3$ since only the
$\pi^{0}\pi^{0}$-channel is considered) is marked on the $y$-axis and the error bars are
indicated by a gray band. In the case of the two resonances below $1$ GeV the
data was taken from the dispersive analysis of Ref. \cite{garcia}, while
for the other two resonances the values from the PDG \cite{pdg} were used. Notice that the PDG value $(350\pm150)$ MeV refers to the full decay width of
$f_{0}(1370).$ We are thus making the simplifying assumption that the $\pi\pi
$-decay mode is dominant (see for instance the recent study in Ref. \cite{stanislaus} and
references therein). This choice represents an upper limit: for smaller $\pi\pi$
branching ratios, the effect of the triangle diagram will be smaller.

The calculation for the resonance $f_{0}(500)$ was done twice: once with a cutoff
$\Lambda\sim10^{5}/2$ GeV (\emph{i.e.}, the practical limit $\Lambda\rightarrow\infty$) and
once with a physical cutoff of $\Lambda\sim1$ GeV in order to demonstrate how the
outcome is influenced by a cutoff value which is typical in hadron physics. As can be seen in Fig. \ref{fig:fig4}, the NLO correction to the decay width is only
important for the resonance $f_{0}(500)$, for which (i) the mass of the decay
products is comparable to $m_{f_{0}(500)}$ \emph{and} (ii) $\Gamma_{f_{0}(500)}^{\text{tree}}\sim
m_{f_{0}(500)}$: in such a case $\Gamma_{\text{NLO}}\sim2\hspace{0.03cm}\Gamma
_{\text{LO}}$, thus the value of the coupling $g$ corresponds roughly to $g^{\prime\prime}$. However, it should be stressed that the NLO process does not include all the other $\pi\pi$-scattering contributions. In the low-energy regime in which $f_{0}(500)$ appears, those contributions interfere and, in virtue of chiral symmetry (see \emph{e.g.} Ref. \cite{garcia}), the full NLO result is expected to be smaller than what our simple model suggests.

For $f_{0}(980)$, $f_{0}(1370)$ and $f_{0}(1500)$ the NLO differs in a small amount with respect to the full result. (Since $f_{0}(1370)$ is quite broad, the NLO correction gives a small but non-negligible contribution to the width. This is so because the ratio $\Gamma_{f_{0}(1370)}^{\text{tree}}/m_{f_{0}(1370)}$ is already in the range of $\sim0.1$. However, the large error bars do not allow to distinguish the NL and the NLO decay widths.) We also observe that a finite cutoff parameter $\Lambda
\sim1$ GeV would affect the decays of $f_{0}(980)$,
$f_{0}(1370)$ and $f_{0}(1500)$ only marginally, since the outcome curve would lie somewhere in between $\Gamma_{\text{LO}}$ and $\Gamma_{\text{NLO}}$.

\begin{figure}
\hspace{-3.0cm}\includegraphics[scale=0.9]{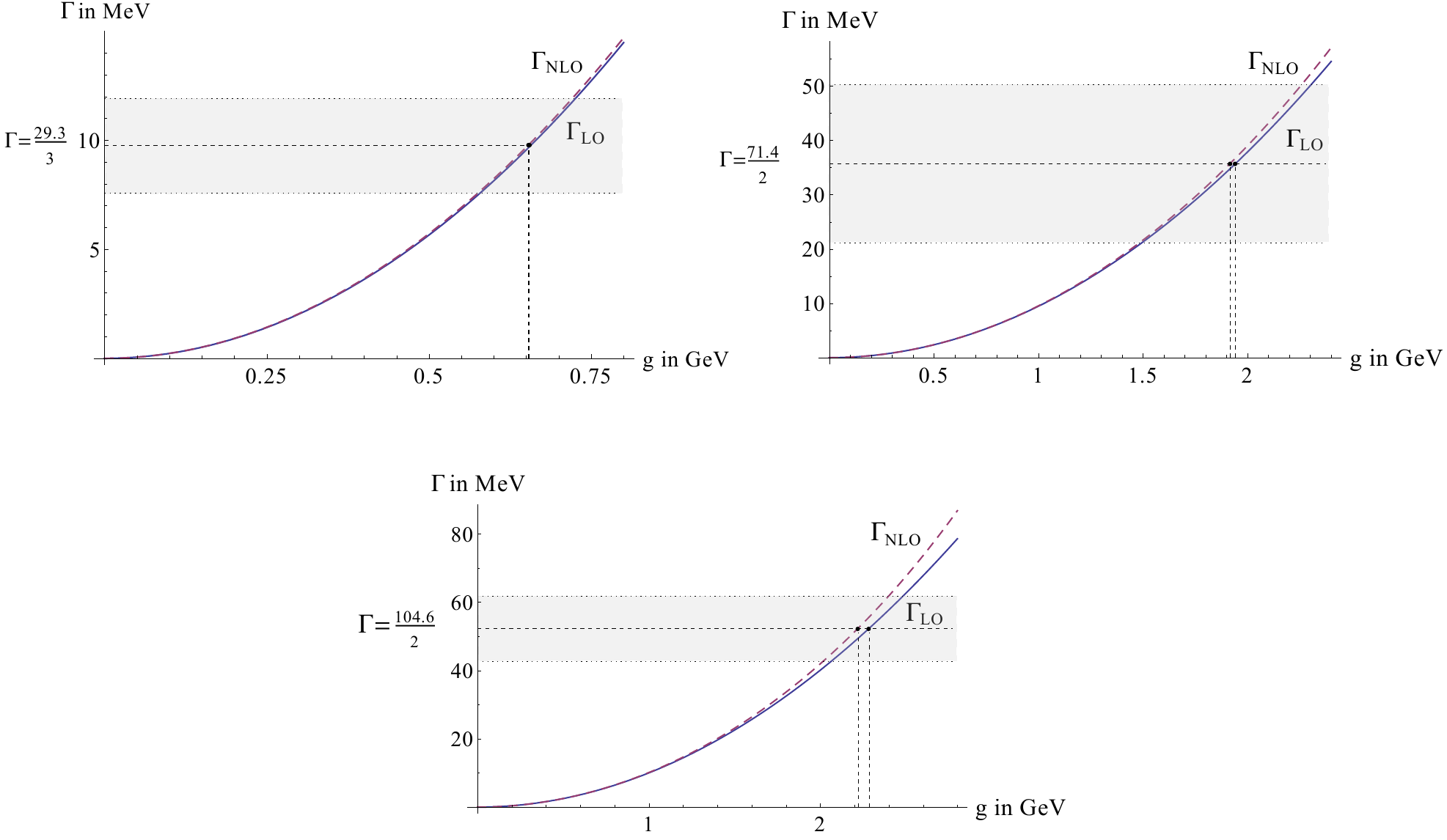}
\caption{Decay widths of $f_{0}(1710)$ going into pions (upper left) and into kaons (upper
right), and $a_{0}(1450)$ into kaons (lower middle). The
`known' value (taken from the PDG \cite{pdg}) of the width is
marked on the $y$-axis and error bars are indicated by a gray band. Note that here $\Gamma=\Gamma_{f_{0}\rightarrow\pi^{0}\pi^{0}}=\Gamma_{f_{0}\rightarrow\pi\pi}/3$ and $\Gamma=\Gamma_{f_{0}/a_{0}\rightarrow K^{+}K^{-}}=\Gamma_{f_{0}/a_{0}\rightarrow K^{0}\bar{K}^{0}}=\Gamma_{f_{0}/a_{0}\rightarrow KK}/2$. The vertical lines correspond to the determination of the coupling constant at LO and NLO, respectively.}
\label{fig:fig5}
\end{figure}

Furthermore, we show in Fig. \ref{fig:fig5} analogous plots for $f_{0}(1710)$ going into pions and kaons, and for $a_{0}(1450)$ going into kaons. The experimental value
of the width (which is $\Gamma_{f_{0}\rightarrow\pi\pi}/3$ and $\Gamma_{f_{0}/a_{0}\rightarrow KK}/2$, respectively) is marked on the $y$-axis and the error bars are again
indicated by a gray band. Here, data was taken from the PDG only \cite{pdg}. By looking at the plots it becomes pretty clear that the NLO correction has only a small influence. Notice that there is a subtle difference in the calculation of the kaon-kaon decay due to the fact that the kaons are distinguishable particles: the factor of $2$ in the LO expression of the amplitude in Eq. (\ref{eq:tree_amplitude_phiphi}), the factor of 8 in the NLO expression in Eq. (\ref{eq:3amplitude}), as well as the symmetry factor $1/2$ in Eq. (\ref{eq:tree_width_phiphi}) are replaced by the unity.

As a next step, we check if the coupling constants obtained in Fig. \ref{fig:fig4} and Fig. \ref{fig:fig5} are compatible with the ones chiral approaches deliver. In Ref. \cite{dick} the resonances $f_{0}(1370)$, $f_{0}(1500)$ and $a_{0}(1450)$ were studied. The various terms in the amplitude for a given channel (\emph{i.e.}, terms with and without derivatives) can be summarized in a unique effective coupling constant for that channel. One gets $\sim 1.7$ GeV for $f_{0}(1370)$ into pions, $\sim 0.7$ GeV for $f_{0}(1500)$ into pions, and $2.28$ GeV for $a_{0}(1450)$ into kaons. Also, the values of the effective couplings of $f_{0}(1710)$ were determined in Ref. \cite{stanislaus} to be $0.64$ GeV for pions and $1.88$ GeV for kaons.

All those values are well compatible with the values shown in Fig. \ref{fig:fig4} and Fig. \ref{fig:fig5}. Similar comments can be made concerning the resonances below $1$ GeV by regarding at the works done in Refs. \cite{giacosaSpectral,e38}. These considerations confirm also that the ranges of couplings studied in Fig. \ref{fig:fig3} were realistic.

\newpage

\clearpage

\section{Conclusions}
\label{seq:conc}
In this work we investigated the importance of the (usually neglected)
triangle-shaped NLO contribution to two-body hadronic decays of the form $S\rightarrow\phi\phi$, see Fig. \ref{fig:fig2}. To this end, the NLO diagram was calculated analytically and numerically in the framework of a quantum field theory involving scalar fields without derivatives (avoiding unnecessary complications due to spin and/or derivatives).

We studied different cases where the mass $m_{\phi}$ of the decay products
varied from nearly zero up to almost $m_{S}/2$, see Fig. \ref{fig:fig3}. The (finite)
contribution from the triangle diagram to the decay width turned out to be
negligible if the mass of the decaying particle is sufficiently large, a
condition which is usually met in hadron physics. We then moved to physical
decays into pions of the well-known scalar resonances $f_{0}(500)$,
$f_{0}(980)$, $f_{0}(1370)$, $f_{0}(1500)$ and $f_{0}(1710)$, as well as to the kaonic channels of $f_{0}(1710)$ and of $a_{0}(1450)$. The outcome was, too, very
clear, see Fig. \ref{fig:fig4} and Fig. \ref{fig:fig5}: the NLO correction gave only a relevant contribution for
$f_{0}(500)$, obviously because $\Gamma_{f_{0}(500)}^{\text{tree}}\sim m_{f_{0}(500)}$. The
resonance $f_{0}(500)$ is however an `extreme example': in all other cases the
correction is much smaller.

\begin{figure}[h]
\centering
\hspace{0.3cm}
\includegraphics[scale=0.71]{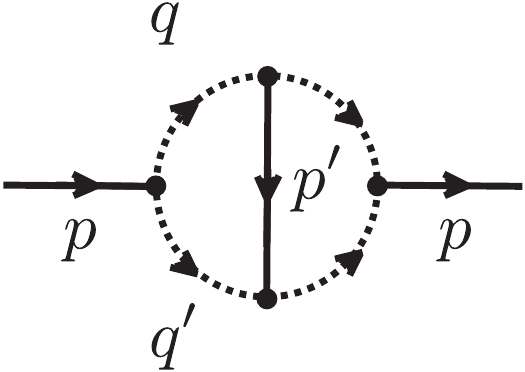}
\caption{NLO correction to the self-energy of the state $S$.}
\label{fig:fig6}
\end{figure}

Within the perturbative framework we have studied in this work, the unstable
state $S$ exchanged in the triangle diagram (Fig. \ref{fig:fig2}) has been considered as stable (\emph{i.e.}, the free propagator was used). Indeed, when considering the fact
that $S$ has a finite width, the contribution from the triangle diagram would be
even smaller. However, the correct way of going beyond the present study is the following:
Besides the one-loop diagram in Fig. \ref{fig:fig1}b, one should perform the resummation of the self-energy of the unstable state $S$ also by incorporating the NLO correction depicted in Fig. \ref{fig:fig6}. Such a study is certainly
nontrivial because the full propagator of $S$ enters here -- one is left with a typical problem of the Bethe--Salpeter type,
see \emph{e.g.} Ref. \cite{alkofer} and references therein. Quite interestingly, the results of
our work show that the modifications coming from such a computation are likely
to be in most cases negligible. Thus, the tree-level results or at most a
description using the (resummed) one-loop propagator of an unstable state, give(s) a
good description of unstable hadronic states.

Other future studies are possible by considering different forms of the
Lagrangian, including derivative interactions, particles with higher spin,
fermionic fields (\emph{i.e.}, baryons), three-body decays and unstable states which
decay in more than one channel.

\section*{Acknowledgements}

The authors thank G. Pagliara, D. H. Rischke and J. Wambach for useful discussions. T. W.
acknowledges financial support from HGS-HIRe, F\&E GSI/GU and HIC for FAIR Frankfurt.

\bigskip

\bigskip

{\small \bigskip}

\end{document}